# Memristor-based Circuits for Performing Basic Arithmetic Operations

Farnood Merrikh-Bayat[a], Saeed Bagheri Shouraki[a]

[a]Department of Electrical Engineering, Sharif University of Technology, Azadi Avenue, Tehran, Iran

**Abstract**

In almost all of the currently working circuits, especially in analog circuits implementing signal processing applications, basic arithmetic operations such as multiplication, addition, subtraction and division are performed on values which are represented by voltages or currents. However, in this paper, we propose a new and simple method for performing analog arithmetic operations which in this scheme, signals are represented and stored through a memristance of the newly found circuit element, i.e. memristor, instead of voltage or current. Some of these operators such as divider and multiplier are much simpler and faster than their equivalent voltage-based circuits and they require less chip area. In addition, a new circuit is designed for programming the memristance of the memristor with predetermined analog value. Presented simulation results demonstrate the effectiveness and the accuracy of the proposed circuits.

## 1. Introduction

Publication of a paper in Nature [1] by Hewlett Packard (HP) labs in May 1, 2008, which announced a first experimental realization of the memristor whose existence was predicted in 1971 by leon chua [2] has caused an extraordinary increased interest in this passive element. In addition to nonvolatile memory applications, it is predicted that memristor can have significant role in designing analog circuits [3] or in human brain emulation [4]. In almost all of the currently working circuits and systems, arithmetic operations are performed on digital operands and this is because of the fact that storing analog values in analog circuits is difficult. It is evident that performing arithmetic operations in analog circuits is too much faster than that in their equivalent digital circuits. Recently, Laiho and Lehtonen proposed memristor-based circuits for performing arithmetic operations which was too much complicated [5]. In this paper, we describe how basic arithmetic operations can be performed on operands which are represented by the memristance of the memristors instead of voltage or current.

The paper is organized as follows. In Section 2, we describe the characteristics of the fourth fundamental circuit element, *i.e.* memristor, and its application in storing analog values. Section 3 explains a procedure of programming memristors with predetermined analog values and performing arithmetic operations on these programmed memristors. Finally, a few experimental results are presented in Section 4, before conclusions in Section 5.

---

\* F. Merrikh-Bayat. Tel.: +98-21-66163657.
   *E-mail address*: f_merrikhbayat@ee.sharif.edu.

## 2. Interesting characteristics of memristor

Memristor is an electrically switchable semiconductor thin film sandwiched between two metal contacts with a total length of $D$ and consists of doped and un-doped regions which its physical structure with its equivalent circuit model is shown in Fig. 1 [1]. The internal state variable $w$ determines the length of doped region with low resistance against un-doped region with high resistivity. This internal state variable and consequently the total resistivity of the device can be changed by applying external voltage bias $V(t)$. This means that passing current from memristor in one direction will increase the resistance while changing the direction of the applied current will decrease its memristance [2]. On the other hand, it is obvious that in this element, passing current in one direction for longer period of time will change the resistance of the memristor more.

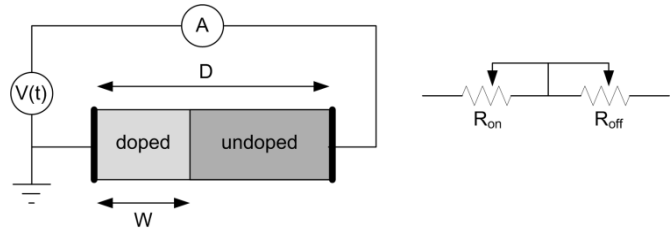

Fig. 1. Structure of the memristor reported by HP researchers and its equivalent circuit model.

As a result, memristor is nothing else than the analog variable resistor which its resistance can be adjusted by changing the direction and duration of the applied voltage or current. Therefore, memristor can be used as a storage device in which analog values can be stored as impedance instead of voltage.

## 3. Basic arithmetic operations based on memristor

For performing any arithmetic operation such as addition, subtraction, multiplication or division, at first, two operands should be represented by some ways. In almost all of currently working circuits, signal values are represented by voltage or current. However, as explained in previous section, analog values can be represented by the memristance of the memristor as well. Figure 2 shows the typical circuit that can be used for adjusting the memristance of one memristor to the predetermined input value, *i.e.* $V_{in}$.

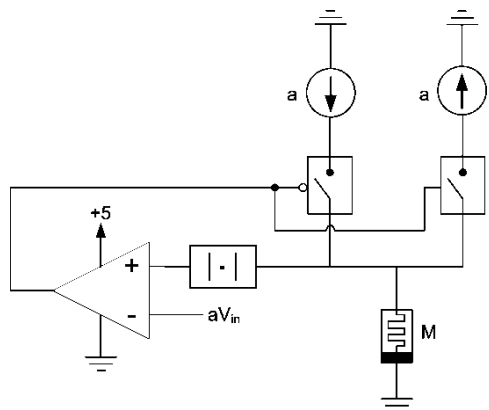

Fig. 2. Typical circuit for adjusting the memristance of the memristor with the predetermined value, $V_{in}$.

In this figure, the coefficient $a$ is considered to make the dropping voltage across the memristor to be meaningful and reasonable. The absolute value of the voltage dropped across the memristor at any time will be $aM$. If $aM$ be lower than $aV_{in}$, the output of the opamp will be at its lowest value, *i.e.* 0 volt, which will cause the left current source to derive the memristor. Passing current from the memristor in this direction will increase its memristance. On the other hand, if $aM$ be higher than $aV_{in}$, the output of the opamp will be at its highest value, *i.e.* 5 volt, which will cause the left current source to derive the memristor. Passing current from the memristor in this direction will decrease its memristance. As a result, final value of the voltage which drops across the memristor, *i.e* $aM$, will be equal to $aV_{in}$ and therefore by this way, the memristance of the memristor will be set to $V_{in}$. Now, this adjusted memristor can be used as an operand for performing arithmetic operations.

Basic arithmetic operations are addition, subtraction, multiplication and division. Performing these operations on the values stored as a memristance of memristors can be simply done by using the circuits shown in Fig. 3. Total memristance of two memristors connected in series is $M_1 + M_2$ which is the addition operation (Fig. 3(a)). Any subtraction, such as $M_1 - M_2$, can be written as $M_1 + (-M_2)$. This means that for doing subtraction, memristor $M_1$ should be connected in series with another memristor which its memristance is $-M_2$. Negative impedance converters such as the one used in Fig. 3(b) can be utilized for creating $-M_2$ from the memristance $M_2$. Figure 3(c) shows a simple opamp-based inverting amplifier which intrinsically is a memristance divider. The output vltage of this circuit is $-(M_2/M_1)V_i$. Setting $V_i = -1$ will cause the output to be equal to $M_2/M_1$. In this case, this circuits acts as a simple divider. Finally, Fig. 3(d) shows the circuit which can be used as a multiplier. By applying standard opamp circuit analysis techniques, output voltage of the circuit is $M_1 \times M_2$. Note that in all of the described circuits, input voltage or current maybe used for scaling the output voltage for avoiding opamps from being saturated.

It is worth to mention that in all of the above circuits, we have assumed that the amplitude or the duration of the applied input voltage or current is too small to change the memristance of the memristors. Consequently, for analyzing of these circuits, a memristor can be considered as a simple resistor.

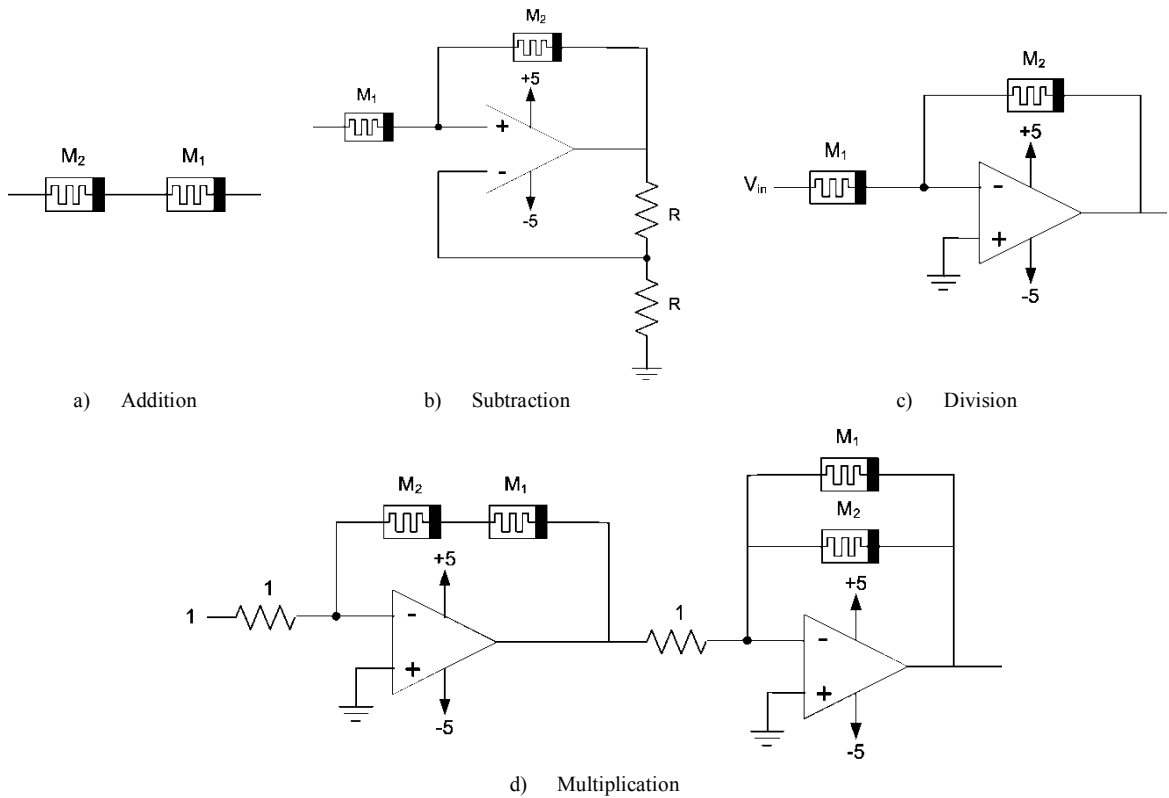

Fig. 3. Typical circuits which can perform basic arithmetic operations.

Some of the main benefits that can be obtained by using these memristance-based arithmetic operators instead of their equivalent conventional voltage or current-based circuits can be summarized as below. First of all, it should be noted that it is predicted that memristor may have a significant rule in a hardware implementation of human brain (for example by emulating neurons and synapses). Therefore, it seems that using memristor-based arithmetic operations in these kinds of circuits are unavoidable. Second, some of these circuits such as divider are so much efficient than their equivalent voltage-based circuits. Finally, values stored in memristors will theoretically remain fixed over an infinite period of time.

## 4. Experimental results

In this section, we want to use the circuit shown in Fig. 3(c) for dividing 520 by 416 which the result should be equal to 1.25. Figures 4(a) and 4(b) demonstrate the results of adjusting memristances of two memristors to 520Ω and 416Ω respectively. Note that these figures show the voltage dropped across the memristors in Fig. 2 during the adjustment procedure while the parameter $a$ is 0.01. These programmed memristors are then used as $M_1$ and $M_2$ in the circuit shown in Fig. 3(c) and the output voltage of this circuit is presented in Fig. 4(c) while the input voltage is a narrow square pulse with the amplitude of -1 volt. It is evident that this circuit can be used as a simple and precise divider.

## 5. Conclusion

In this paper, we proposed a new method for performing basic arithmetic operations on operands which are represented by the memristance of the memristors. A new circuit is suggested for the programming of the memristance of the memristor with the predetermined value. Proposed circuits are simple and some of them are much efficient than that their equivalent voltage or current-based circuits.

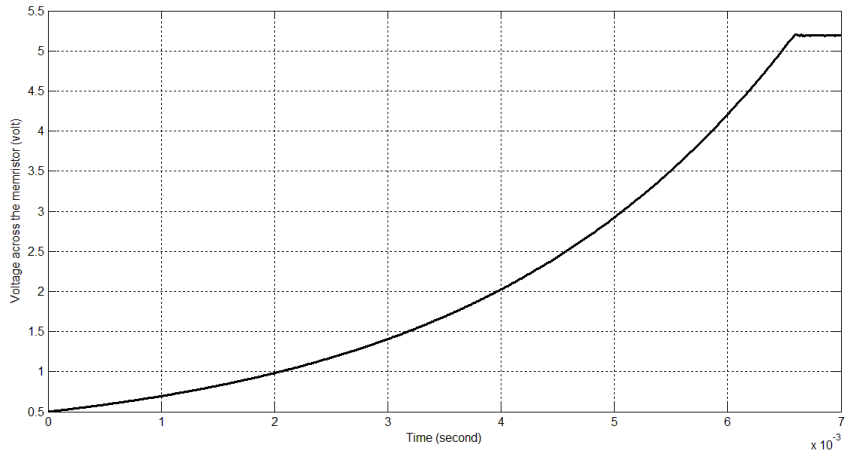

a)

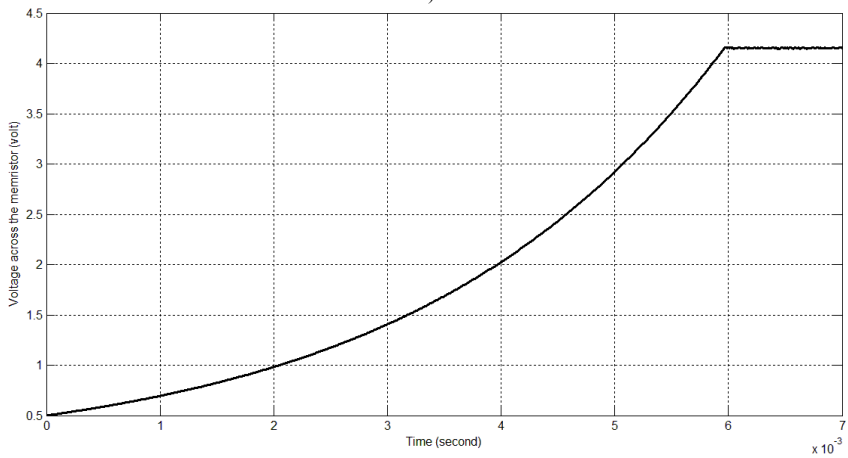

b)

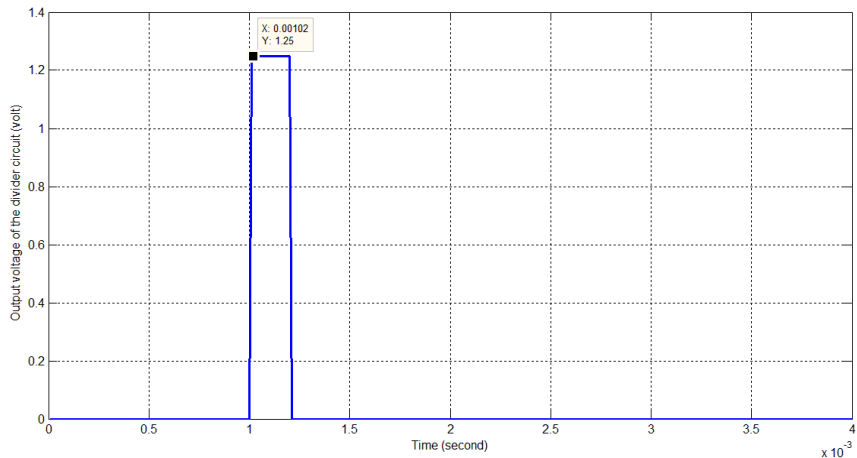

c)

Fig. 4. a) Adjusting the memristance of the first memristor to 520Ω. b) Adjusting the memristance of the second memristor to 416Ω. c) Output voltage of the circuit shown in Fig. 3(c) while these two programmed memristors are used in it as $M_1$ and $M_2$.